# CAN THE SOLAR ANGULAR MOMEMTUM PROBLEM BE SOLVED WITHOUT INTERNAL GRAVITY WAVES?

*This work is dedicated to the memory of Italo Mazzitelli*

V.M. Canuto [1,2]

[1]NASA, Goddard Institute for Space Studies, 2880 Broadway, NY, NY, 10025
[2]Dept. of Applied of Phys. and Math., Columbia University, NY, NY, 10027.

**Abstract.** The solar radiative zone spin down problem is treated without internal gravity waves, IGW. The shear-only model of the Reynolds stresses adopted thus far in all calculations, is substituted by a shear + vorticity model. The latter component is shown to play a role analogous to the IGW and with a similar time scale of about $\approx 10^7$ years.

## I. The problem

Helio-seismological data have offered the following new picture of the solar interior: a stably stratified radiative zone RZ of about 70% of the solar radius characterized by unform rotation, and an upper convective zone CZ of about 30% of the solar radius characterized by differential rotation. Fig.7 of the comprehensive review article by Thompson et al. (2003, T4), is a clear illustration of the above structure. The main goal of several of the past studies cited in T4 was that of suggesting physical processes capable of extracting angular momentum from the RZ so as to drive it toward the state of the observed uniform rotation in a time scale of the order of $10^7$ years. The most frequently cited process that accomplishes the desired result is the suggestion of IGW (internal gravity waves) reviewed in sec 7.3 of T4. This paper is a contribution to the same problem by suggesting a heretofore untapped mechanism that may prove relevant to the spin-down problem. Anticipating the details presented below, the key new feature is the introduction of a *verticity-based Reynolds stress* in addition to the traditional shear driven model. As shown in sec 5.1 of T4, the time evolution of the angular-momentum density J due to meridional (MC) and Reynolds stresses (RS) is given by:

$$J = \rho r^2 \Omega \sin^2\theta \quad, \quad \frac{\partial J}{\partial t} = -\nabla \cdot (\mathbf{F}_{MC} + \mathbf{F}_{RS}) \tag{1}$$

where:

$$\mathbf{F}_{MC} = J(U_\theta \hat{\boldsymbol{\theta}} + U_r \hat{\mathbf{r}}), \quad \mathbf{F}_{RS} = r\sin\theta (<\rho u_\theta u_\varphi> \boldsymbol{\theta} + <\rho u_r u_\varphi> \hat{\mathbf{r}}) \tag{2}$$

in which capital and low character symbols indicate mean and fluctuating fields. Specifically, one has the equation:

$$\Gamma \frac{\partial J}{\partial t} = -\frac{1}{r^2}\frac{\partial}{\partial r}r^3(b_{r\varphi} + U_r U_\varphi) - \Gamma^{-2}\frac{\partial}{\partial \theta}\Gamma^2(b_{\theta\varphi} + U_\theta U_\varphi) \tag{3}$$

where:

$$b_{ij} = \overline{u_i u_j} - \frac{2K}{3}\delta_{ij}, \quad K = \frac{1}{2}\overline{u_i u_i} \tag{4}$$

are the traceless Reynolds stresses, $R_{ij} \equiv \overline{u_i u_j}$.

## 2. The shear model for the Reynolds stress

We begin by considering the simplest, shear driven RS:

$$R_{ij} = -2K_m S_{ij}, \quad S_{ij} = \frac{1}{2}(U_{i,j} + U_{j,i}) \tag{5}$$

In spherical coordinates we have:

$$R_{r\varphi} = -K_m r\sin\theta \frac{\partial \Omega}{\partial r}, \quad R_{\theta\varphi} = -K_m \sin\theta \frac{\partial \Omega}{\partial \theta} \tag{6}$$

In the solar case, $\Omega$ omega increases toward the equator and thus $\sin\theta \frac{\partial \Omega}{\partial \theta} > 0$ in the northern hemisphere and $<0$ in the southern hemisphere, so that the second relation in (6) implies that:

$$R_{\theta\varphi} < 0 \text{ (N)} \quad, \quad R_{\theta\varphi} > 0 \text{ (S)} \tag{7}$$

while the observational data cited in sec.6 of Canuto et al. (1994) show the opposite behavior.

To remedy the situation, in sec.8.1 of the same work, the above authors studied five combinations of shear, vorticity and buoyancy. Only the last combination yielded results that reproduced the data, as shown in fig.1,2 of that paper. Though unrelated to the interior RZ of interest here, the unexpected emergence of vorticity in lieu of the commonly accepted shear, was a telling surprise that partially motivated our search for its role in the spin-down of the RZ. Using the first of (6) in (3) without the mean fields, gives rise to the following time evolution of J:

$$\frac{\partial J}{\partial t} = \frac{1}{r^2}\frac{\partial}{\partial r}(K_m r^4 \frac{\partial \Omega}{\partial r}) \tag{8}$$

which T4 wrote "*it predicts rotation of the solar interior at a rate several times higher than the surface rate, in stark disagreement with helio data of nearly uniform rotation*".

## 3. The IGW suggested solution

As discussed in Talon and Charbonnel (2005), the suggestion was made to add to Eq.(8) the contribution due to internal gravity waves (IGW):

$$\frac{\partial J}{\partial t} = \frac{1}{r^2}\frac{\partial}{\partial r}(K_m r^4 \frac{\partial \Omega}{\partial r}) - \frac{3}{8\pi\rho}\frac{1}{r^2}\frac{\partial}{\partial r}L_{IGW} \tag{9}$$

where the magnitude of the IGW luminosity is given as $10^{29}$ ergs$^{-1}$ or 0.004 percent of the luminosity at the base of the CZ; from (9), the time scale on which IGW operate is about $10^7$ yrs.

## 4. Vorticity in the RZ

Including shear and vorticity means that the Reynolds stress in the J equation (3) is now:

$$R_{r\varphi} = -2(K_m S_{r\varphi} + xK_m V_{r\varphi}) \tag{10}$$

where $x = (\tau N)^2$ is the dimensionless dynamical time scale, N is the Brunt-Vaisala frequency and:

$$S_{r\varphi} = \frac{1}{2}r\sin\theta\frac{\partial \Omega}{\partial r} \quad , \quad V_{r\varphi} = -\frac{1}{2r}\sin\theta\frac{\partial r^2\Omega}{\partial r}$$

(11)

which yield the new form of the J equation:

$$\frac{\partial J}{\partial t} = r^{-2}\frac{\partial}{\partial r}(r^4 K_m \frac{\partial \Omega}{\partial r}) - r^{-2}\frac{\partial}{\partial r}(Xr^2 K_m \frac{\partial r^2\Omega}{\partial r}) \tag{12}$$

A few remarks may be useful. *First*, the sign of the vorticity term in (11) is opposite to that of shear, *second*, even in the constant rotation case, the vorticity term does not vanish but becomes:

$$-\Gamma\Omega, \quad \Gamma=\sin\theta \tag{13}$$

that reproduces the so-called $\Lambda-$effect introduced heuristically by Rudiger (1989); *third* and most relevant to the present case, the vorticity term contains the new ingredient:

$$X=(\tau N)^2, \quad X>0 \text{ (RZ)}, \quad X<0 \text{ (CZ)} \tag{14}$$

*It is only because of the second relation in (14) that the sign of the vorticity term in (12) is negative as it is in the IGW case of Eq.(9).*

## 5. Time scale

In the new vorticity term in (12), the corresponding time scale is given by:

$$t \approx \frac{r^2}{K_m} \frac{1}{X} \approx \frac{r^2 N^2}{\varepsilon} \frac{1}{\Gamma_m G_m Ri}, \quad G_m \equiv (\tau\Sigma)^2 \tag{15}$$

where $\Sigma$ is he mean shear and use was made of the well-known relation:

$$K_m = \Gamma_m \frac{\varepsilon}{N^2} \tag{16}$$

Using the following data from Talon and Charbonnel (2005):

$$r=3.5*10^9 \text{cm}, \quad \Omega=2*10^{-6}\text{s}^{-1}, \quad \varepsilon=3*10^{-5}\text{cm}^2\text{s}^{-3}, \quad N^2=10^{-7}\text{s}^{-2}, \quad \varepsilon N^{-2}=300\text{cm}^2\text{s}^{-1} \tag{17}$$

Eq. (15) becomes

$$t \approx F(Ri)10^7 \text{yrs}, \quad F(Ri)=\frac{4}{3}\frac{10^3}{G_m\Gamma_m Ri} \tag{18}$$

where the function $G_m$ vs. Ri is given in Fig.3a of C10, and the mixing efficiency $\Gamma_m$ vs. Ri is shown in Fig.2a. For Ri=1, $G_m=10^2$, $\Gamma_m=1$, it follows that F(Ri) is of O(1) and thus:

$$t \approx 10^7 \text{yrs} \tag{19}$$

which is the same as the one of the IGW. The function F(Ri) vs. Ri is plotted in Fig.1.

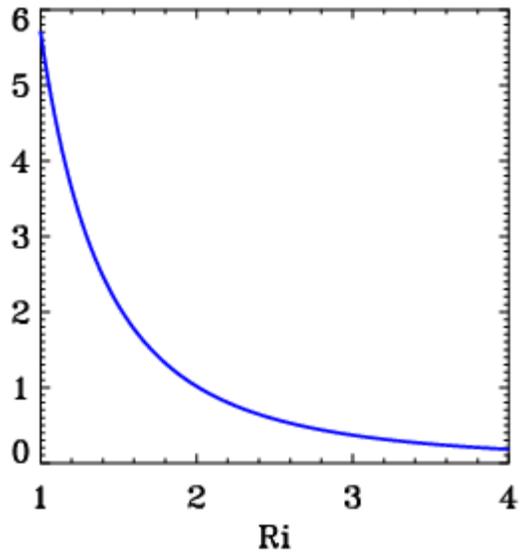

Fig.1. The function F(Ri) defined in Eq.(18) vs Ri.

## 6. About $Ri_{cr}$

For many years, it was believed there is a critical Ri above which there is no turbulent mixing.

In Canuto et al. (2008), it was shown there is no $Ri_{cr}$ and Fig.3f of C10 cites the relevant data.

## 7. Shear driven model: an anticipated failure?

Some after-thought may be in order concerning the failure of the shear model. The fault lies at the heart of such a model since it assumes that the principal axes of the Reynolds stresses representing turbulence, be aligned with those of the tensor $R_{ij}$ representing the mean flow. This is true only in the particular case of *pure strain which* is far from being the case in the highly turbulent solar interior. The demise of the pure-strain model can therefore be viewed as a built-in shortcoming.

## 8. Conclusions

The above computations shows that there is an alternative to IGW which is unavoidable since the presence of vorticity is not an assumption but a logical extension of the assumed shear-only model which had no physical justification other than simplicity. All Reynolds stress models derived in C10, for example Eq.A.6, exhibit such a term which, perhaps not coincidentally, was also instrumental in explaining the data discussed in sec.2.

Acknowledgement

I thank Juri Toomre for his comments.

**Data Policy**

This study does not utilize or create data.